\documentclass[hyper,11pt,paper,superscriptaddress,nofootinbib,prd]{article}
\pdfoutput=1
\usepackage{jheppub}

\usepackage[latin1]{inputenc}
\usepackage{amsmath}
\usepackage{amsfonts}
\usepackage{amssymb}
\usepackage{graphicx}
\usepackage{xcolor}
\usepackage{mathtools}
\usepackage{simplewick}
\usepackage{hyperref}
\usepackage{multirow}

\newcommand{\exv}[1]{\left\langle #1 \right\rangle}

\newcommand{\specialcell}[2][c]{%
	\begin{tabular}[#1]{@{}c@{}}#2\end{tabular}}

\begin{document}

\title{Probing emergent geometry through phase transitions in free vector and matrix models}

\author{Irene Amado$^1$, Bo Sundborg$^1$, Larus Thorlacius$^{1,2}$ and Nico Wintergerst$^1$}

\affiliation{$^1$The Oskar Klein Centre for Cosmoparticle Physics, Department of Physics, Stockholm University, AlbaNova, 106 91 Stockholm, Sweden}

\affiliation{$^2$University of Iceland, Science Institute, Dunhaga 3, 107 Reykjavik, Iceland}

\emailAdd{irene.amado@fysik.su.se}
\emailAdd{bo@fysik.su.se}
\emailAdd{larus.thorlacius@fysik.su.se}
\emailAdd{nico.wintergerst@fysik.su.se}

\abstract{
Boundary correlation functions provide insight into 
the emergence of an effective geometry in higher spin 
gravity duals of $O(N)$ or $U(N)$ symmetric field theories. On a compact manifold, 
the singlet constraint leads to nontrivial dynamics at finite temperature and 
large $N$ phase transitions even at vanishing 't Hooft coupling.
At low temperature, the leading behavior of boundary two-point functions
is consistent with propagation through a bulk thermal anti de Sitter space. 
Above the phase transition, the two-point function shows
significant departure from thermal AdS space and the emergence 
of localized black hole like objects in the bulk. 
In adjoint models, these objects appear at length scales of order of the 
AdS radius, consistent with a Hawking-Page transition, but in vector models 
they are parametrically larger than the AdS scale. In low dimensions, we find another 
crossover at large distances beyond which the correlation
function again takes a thermal AdS form, albeit with a temperature dependent 
normalization factor.
}
    
\maketitle

\section{Introduction}

Under gauge/gravity duality, gravity and spacetime are viewed as emergent phenomena 
that arise from quantum dynamics and entanglement in a dual field theory. Based on this,
gravitational methods have been used to address a range of problems in strongly coupled field 
theories, where the emergent spacetime is well described by classical field equations, but
gauge/gravity duality also has important ramifications for quantum gravity. In fact, if such a 
duality holds in Nature, it obviates the need for a fundamental quantum theory of the 
gravitational interaction and suggests that thorny issues in gravitational theory, such as 
the resolution of spacetime singularities, may ultimately be explained in terms of a weakly 
coupled quantum field theory.  As a step in that direction, we study simple field theories 
that are believed to have gravitational duals and explore their weak coupling dynamics 
by computing low order correlation functions. 

The simplest suitable field theories are $O(N)$ or $U(N)$ symmetric theories at large $N$, 
which are free except for a singlet constraint on the states. They are conjectured to be
dual to Vasiliev higher spin theory 
\cite{FradkinVasiliev1987b,FradkinVasiliev1987a,KlebanovPolyakov2002,GaberdielGopakumar2011} 
and tensionless string theory in AdS \cite{Haggi-ManiSundborg2000,Sundborg2001}, 
respectively. The singlet constraint is far from innocuous, however. In finite volume it leads 
to non-trivial large $N$ phase transitions \cite{Sundborg2000,ShenkerYin2011}. In the limit of 
infinite string tension, {\it i.e.} semi-classical gravity limit, the corresponding 
confinement-deconfinement phase transition 
has been identified with the Hawking-Page transition involving bulk black 
holes \cite{HawkingPage1983,Witten1998a} and it is an interesting open question to what 
extent the large $N$ phase transitions in the above theories with higher spin symmetries 
involve black hole-like objects as well. So far, there is little concrete evidence for such 
an identification beyond gross thermodynamic features. The results are also non-uniform. 
There are recent indications that there is no Hawking-Page transition at AdS scale in 
three-dimensional higher spin gravity \cite{BanerjeeCastroHellermanHijanoLepage-JutierMaloneyShenker2013,GaberdielGopakumarRangamani2014} 
due to the presence of light states that smooth the transition. 
Likewise, there appear to be no black holes in thermal equilibrium at the AdS scale in 
pure higher spin theory in four dimensions but there is a phase transition at a much higher 
temperature (Planck scale) in that theory \cite{ShenkerYin2011}. In contrast, AdS scale 
objects do appear to exist in the theory dual to tensionless strings. In this case, the  't~Hooft 
coupling can be taken to be non-zero leading to tensile strings and eventually to the infinite tension 
gravity limit at strong 't~Hooft coupling where the role of black holes is well understood. 

Some insight into the topology of the Euclidean background can be gained by considering 
the expectation value of the square of a Polyakov loop in the $U(N)$ matrix 
model \cite{Furuuchi2005}. 
In particular, a non-vanishing expectation value indicates a contractible thermal circle 
in the dual geometry corresponding to a deconfined phase. 
For higher spin theory, partial geometric insights from boundary theory thermodynamics have 
been obtained recently by considering the high temperature phase far above the phase transition 
and identifying evanescent excitations behaving as if traveling along an event horizon
\cite{JevickiYoon2016,JevickiSuzuki2016}.  
This is intriguing and promising, and it could be illuminating to analyze the problem using 
complementary methods, especially if they can bring us closer to the phase transition.  

In the present paper, we extend the thermodynamic analysis by considering 
boundary correlation functions at thermal equilibrium in a vector model and a 
matrix model with $U(N)$ symmetry.\footnote{Our results can easily be extended to the $O(N)$ 
vector model considered in \cite{KlebanovPolyakov2002}.}
The correlation functions involve gauge invariant observables in the 
boundary theories and we study their spatial dependence as a function of temperature\footnote{For a recent review of the higher spin / vector model duality, including many results on zero-temperature correlation functions, see \cite{Giombi2016}.}, 
identifying leading behaviors across different thermodynamic phases in these models. 
We will be working in a free field limit on the field theory side and looking to interpret
the resulting correlation functions in terms of an emergent spacetime geometry. 
It is not clear {\it a priori\/} that this will work in the absence of a large 't~Hooft coupling 
but our results are nevertheless suggestive 
of an AdS geometry emerging at low temperatures in both models and signs of more 
elaborate geometric structures emerge above their respective phase transitions.   

We calculate boundary correlation functions at finite temperature in a large $N$ 
expansion. We allow for an arbitrary number of flavours $N_{f}$ in the calculations, 
but $N_{f}$ is always kept finite, and as a result the flavor dependence is trivial. 
In contrast, we find interesting dependence on spacetime dimension, which we 
parametrize by the boundary dimension $d$. 
We note that $d=3$ is the natural dimension for the vector 
model (dual to massless higher spins in four-dimensional spacetime) while 
$d=4$ is the natural dimension for the matrix model (dual to tensionless strings in
five dimensions). In these dimensions there are nearby interacting models that are 
conformal and allow for non-trivial AdS/CFT studies. We will nevertheless consider 
free theories in higher dimensions, viewing them as useful toy models, but keeping in 
mind that in this case the effect of turning on interactions is much more uncertain.
We find clear evidence of spatial structure in the putative dual spacetime to the large~$N$ 
high temperature phases in all these models. The objects we find either appear at the 
AdS scale (matrix model) or at much larger length scales (vector model). We also find 
a range of temperatures above the phase transition in each model, where the  
thermal objects appear to have a central core exhibiting a novel structure
when probed by boundary-to-boundary two point functions at large operator separation.

The paper is structured as follows. In Section~\ref{sec:saddle} we review standard 
techniques for imposing singlet constraints and exposing the thermal phase 
transitions in the large~$N$ limit on $S^{d-1} \times S^1$ for both vector and adjoint 
(matrix) models. We apply the large $N$ technology to obtain Green's functions 
in Section~\ref{sec:thermalgreen} and then study two-point correlation functions 
of gauge invariant observables and their behavior across different temperature 
and angular regimes, in Section~\ref{vectormodel_2pt} for the vector model, and in 
Section~\ref{sec: adjointmodel_2pt} for the adjoint model. 
Finally, in Section~\ref{sec:discussion} we summarize and briefly discuss our results. 
In Appendix~\ref{sec:app3d}, explicit expressions are worked out for the 
important special case of the $d=3$ vector model, which is conjectured to be dual to 
four-dimensional Vasiliev higher spin theory. In Appendix~\ref{appendixb} we 
evaluate expectation values of Polyakov loops in both the vector and adjoint models.

\section{Partition function, saddle points and phase transition}\label{sec:saddle}

In this section we provide the necessary background for our analysis, for the most part following 
\cite{AharonyMarsanoMinwallaPapadodimasVan-Raamsdonk2004,ShenkerYin2011}.
We commence by considering the partition function
\begin{equation}
	Z[\beta] = \int {\cal D}A_\mu {\cal D}\phi {\cal D}\phi^\dagger 
	e^{-S[A_\mu, \phi, \phi^\dagger; \beta]}\,,
\end{equation}
where the fields live on a spatial $d{-}1$-sphere and a thermal circle $t\cong t+\beta$, 
leading to $S^{d-1} \times S^1$ as the underlying manifold.\footnote{We set the radius of the 
$S^{d-1}$ to $R=1$ to simplify notation. All lengths in the boundary theory are referred to this 
length scale.}
The action is that of a $U(N)$ gauge theory in the limit of a vanishing gauge coupling.
We consider two cases for the matter sector. Either it consists of $N_f$ free scalar fields in the 
fundamental representation, or of a scalar in the adjoint representation, in both cases with 
a conformal coupling to the background curvature.
As usual, gauge invariance implies Gauss' law constraints. 
Due to the compactness of the $S^{d-1}$, the integral form of these constraints 
imposes vanishing of all charges. 

For this work, we are interested in the free limit of the gauge theory. In this case, the only 
remaining effect of the gauge coupling is to enforce the singlet constraint. The nonzero 
modes of the gauge field decouple entirely, and we will ignore them from this point on. 
However, a nontrivial contribution arises from the zero mode of the gauge field,
\begin{equation}
	\alpha(t) = \frac{1}{Vol(S^{d-1})} \int_{S^{d-1}} A_t\,,
\end{equation}
which we may gauge fix to be a constant in Euclidean time,
\begin{equation}
	\partial_t \alpha = 0\,.
\end{equation}
Separation of $A$ and $\alpha$ in the path integral through a $U(N)$ rotation 
$U_\alpha = e^{i \alpha t}$ leads to the scalar action
\begin{equation}
\label{eq:scal_ac}
	S_\phi = \int -\phi^\dagger\left(D_{t}^2 + \partial_i^2 - \frac{1}{4}(d-2)^2\right)\phi\,,
\end{equation}
with $\phi$ either in the fundamental or the adjoint representation. 
The covariant derivative is given by $D_{t} = \partial_t + i\alpha$ in the vector model and 
by $D_{t} = \partial_t + i[\alpha,\cdot]$ for an adjoint scalar.

Note that in addition to changing the form of the action, the rotation has changed the 
boundary conditions around the thermal circle to twisted ones
\begin{equation}
	{\cal O}(t = 0) = \tilde{\cal O}(t = \beta)\,,
\end{equation}
where $\tilde{\cal O}$ is the operator obtained by applying the $U(N)$-rotation 
$e^{i\alpha\beta}$. This fixes the singlet condition on all field configurations. 
In other words, the integral only runs over gauge invariant field configurations.

Including and evaluating gauge fixing terms and corresponding Fadeev-Popov 
determinants \cite{AharonyMarsanoMinwallaPapadodimasVan-Raamsdonk2004} 
and integrating over the scalar leads to an effective matrix model for $\alpha$, or, 
equivalently, of the $U(N)$ gauge holonomy around the thermal circle, 
$U = e^{i\beta\alpha}$. Its partition function is most conveniently written in terms 
of the eigenvalues $e^{i\lambda_i}$ of the holonomy matrix $U$. The partition function 
reads \cite{Sundborg2000,AharonyMarsanoMinwallaPapadodimasVan-Raamsdonk2004,ShenkerYin2011,GrossWitten1980}
\begin{equation}
	\label{eq:genfun}
	Z[\beta] = \frac{1}{N!} \int \left(\prod_i d\lambda_i\right) 
	\exp\Bigg[  \sum_{i\neq j} \ln\left|\sin\left(\frac{\lambda_i - \lambda_j}{2}\right)\right|
	- \sum_i S_s[\lambda_i] \Bigg] \,.
\end{equation}
$S_s[\lambda]$ is the scalar contribution to the partition sum.  In the vector model it is
given by
\begin{equation}
S_s[\lambda_i] = -2 N_f \sum_{k=1}^{\infty}\frac{1}{k}z_S^d(x^k) \cos(k\lambda_i) \,,
\end{equation}
and in the adjoint model by
\begin{equation}
S_s[\lambda_i] =  -\sum_{k=1}^{\infty} \frac{1}{k} z_S^d(x^k) \sum_ j \cos(k(\lambda_i - \lambda_j))\,.
\end{equation}
The scalar one-particle partition sum on a $d-1$-dimensional sphere is given by
\begin{equation}
	z_S^d(x)=x^{\frac{d}{2} - 1}\frac{1+x}{(1-x)^{d-1}}\,,
\end{equation}
with $x = e^{-\beta}$.

One could now work directly with Eq.~\eqref{eq:genfun} and look for saddle points by varying 
the action with respect to the eigenvalues, as done for example in \cite{ShenkerYin2011}. 
In order to compare the vector and matrix models in the large $N$ limit, however, it proves 
convenient to rewrite the action by defining
\begin{equation}\label{eq:rho_k}
\rho_k = \frac{1}{N}\sum_i \cos k\lambda_i\,,
\end{equation}
where the normalization has been chosen s.t. for a homogeneous distribution of eigenvalues $\rho_0 = 1$. Moreover, note that
\begin{equation}
\ln\left|\sin\left(\frac{\lambda_i - \lambda_j}{2}\right)\right| =  -\log 2 - \sum_{k=1}^\infty \frac{1}{k}\cos(k(\lambda_i - \lambda_j))\,.
\end{equation}
After some straightforward manipulations, and under the assumption of a symmetric distribution, we obtain for the action\footnote{In fact, there is a subtlety when writing the action in this form. As can be easily checked, there is a contribution to the action that arises from $i = j$ in Eq.~\eqref{eq:genfun} and which diverges logarithmically. In the full expression, this contribution is subtracted. We ignore the subtraction term, however, since it has no impact in the large $N$ limit, when the $\rho_k$ can be treated as independent variables.} in the vector model
\begin{equation}
\label{eq:act_fun}
S = N^2\sum_k\frac{\rho_k}{k}\left(\rho_k - 2\frac{N_f}{N}z_S^d(x^k)\right)\,,
\end{equation}
and in the adjoint model
\begin{equation}
\label{eq:act_adj}
S = N^2 \sum_k\frac{1 - z_S^d(x^k)}{k} \rho_k^2\,.
\end{equation}
In the latter case, the $\mathbb{Z}_n$ center symmetry is no longer apparent, due to the 
assumption of a symmetric distribution of eigenvalues. Later, we will remedy this by 
including appropriate factors accounting for the sum over all saddles.

\subsection{Saddles}

For large $N$, the path integral Eq.~\eqref{eq:genfun} can be evaluated in a saddle 
point approximation. To leading order in $1/N$, the saddle point equation is most readily 
solved in the continuum approximation. To this end, we introduce the eigenvalue 
density $\rho(\lambda)$, normalized to
\begin{equation}
\label{eq:distnorm}
\int d\lambda\,\rho(\lambda) = 1\,,
\end{equation}
and obeying the positivity constraint
\begin{equation}
\label{eq:pos}
\rho(\lambda) \geq 0\,.
\end{equation}
The $\rho_k$ then have a straightforward interpretation as the Fourier cosine 
coefficient of the eigenvalue distribution. 

\subsubsection{Vector model} 
In the vector model, it is easy to see that the action Eq.~\eqref{eq:act_fun} is now minimized for 
\begin{equation}
\rho_k = \frac{N_f}{N}z_S^d(x^k)\,.
\end{equation}
Together with the normalization condition Eq.~\eqref{eq:distnorm} this yields \cite{GrossWitten1980}
\begin{equation}
	\label{eq:dist_sum}
	\rho(\lambda) = \frac{1}{2\pi} + \frac{N_f}{N}\sum_{k=1}^{\infty} z_S^d(x^k) \frac{1}{\pi}\cos(k\lambda)\,.
\end{equation}

This solution is valid up to temperatures such that the positivity constraint Eq.~\eqref{eq:pos} is no longer satisfied. This happens when the distribution Eq.~\eqref{eq:dist_sum} vanishes somewhere along the circle and marks the onset of a Gross-Witten type phase transition. In the large $N$ limit, for fixed finite $N_f$, it takes place at very high temperatures $T\sim\sqrt{N}$.

We can find the critical temperature by noting that at high temperatures, $1-x \approx 1/T$ and the scalar partition function becomes
\begin{equation}
\label{eq:zs_ht}
	z_S^d(x^k) \to 2\,\frac{T^{d-1}}{k^{d-1}}\,.
\end{equation}
Moreover, in our conventions, the minimum of the distribution lies at $\lambda = \pm \pi$. Inserting this into Eq.~\eqref{eq:dist_sum} and evaluating the sum leads to
\begin{equation}
\rho(\pi) = \frac{1}{2\pi}-\frac{2}{\pi}\,(1-2^{2-d})\zeta(d-1)\,\frac{N_f}{N}\,T^{d-1}\,,
\end{equation}
which fixes the critical temperature to
\begin{equation}
T_c^{d-1} = \frac{1}{4\left(1-2^{2-d}\right)\zeta(d-1)}\, \frac{N}{N_f}\,.
\end{equation}

At higher temperatures, the eigenvalue distribution vanishes for a finite interval of $\lambda$.
It is described by 
\begin{equation}
\rho(\lambda) = 
\begin{cases}
A_d\left(\frac{T}{T_c}\right)^{d-1}\left(\Re Li_{d-1}(e^{i\lambda}) 
- \Re Li_{d-1}(e^{i\lambda_m})\right)&\text{for } |\lambda| \leq \lambda_m\,,\\
0&\text{else}\,,
\end{cases}
\end{equation}
where we have defined $A_d = \frac{1}{2\pi\left(1-2^{2-d}\right)\, \zeta(d-1)}$, and 
$\lambda_m$ is fixed by the normalization condition Eq.~\eqref{eq:distnorm}. Here,  $ Li_{n}$ 
denotes the $n$-th Polylogarithm and $\Re$ and $\Im$ are the real and imaginary parts, 
respectively. Explicitly, we have
\begin{equation}
\label{eq:norm_d}
2A_d\left(\frac{T}{T_c}\right)^{d-1}\left( \Im Li_{d}(e^{i\lambda_m})-\lambda_m\Re Li_{d-1}(e^{i\lambda_m})\right) = 1\,.
\end{equation}
Analytic solutions for $\lambda_m$ can be found for $d = 3$ and are given in appendix \ref{sec:app3d}. 

Focusing on the scaling behavior far above the transition, $T \gg T_c$,
we expand Eq.~\eqref{eq:norm_d} in powers of $\lambda_m$ and solve the lowest order equation,
\begin{equation}
\lambda_m \to \frac{T_c}{T} \times
\begin{cases}
\sqrt{\frac{2}{\pi A_d}} &\text{for }d=3\,,\\
\left(\frac{3}{2A_d}\right)^{\frac{1}{3}}\log^{-1/3}\frac{T}{T_c}&\text{for }d=4\,,\\
\left(\frac{3}{2\zeta(d-3)A_d}\right)^{\frac{1}{3}}\left(\frac{T_c}{T}\right)^{\frac{d-4}{3}} &\text{for }d\geq 5\,.
\end{cases}
\end{equation}

\subsubsection{Adjoint model} 
The solution in the adjoint case is found in the same way, but the mechanism 
driving the phase transition is somewhat different.
From Eq.~\eqref{eq:act_adj} one observes that the homogeneous distribution, $\rho_{k \geq 1} = 0$, is preferred as long as $z_S^d(x^k)< 1$ for all $k$. At temperatures $T = T_c \sim \mathcal{O}(1)$, $z_S^d(x_c) = 1$, and the first Fourier mode of the distribution becomes gapless. This is the Hagedorn transition, beyond which $\rho_1$ condenses. For higher temperatures, the analysis requires taking into account the positivity constraint and becomes more involved. There, the constraint sets conditions on all $\rho_n$. Nevertheless, it is clear that the distribution vanishes at a single point just above the transition, and subsequently on a finite interval.

An approximate expression for the distribution at $T>T_c$, given in \cite{AharonyMarsanoMinwallaPapadodimasVan-Raamsdonk2004}, reads
\begin{equation}
\rho(\lambda) = 
\begin{cases}
\frac{1}{\pi \sin^2\left(\frac{\lambda_m}{2}\right)}\sqrt{\sin^2\left(\frac{\lambda_m}{2}\right) - \sin^2\left(\frac{\lambda}{2}\right)}\cos\left(\frac{\lambda}{2}\right)&\text{for } |\lambda| \leq \lambda_m\,,\\
0&\text{else}\,,
\end{cases}
\end{equation}
where the maximal angle $\lambda_m$ is fixed by
\begin{equation}
\sin^2\left(\frac{\lambda_m}{2}\right) = 1 - \sqrt{1-\frac{1}{z_S^d(x)}}\,.
\end{equation}
At high temperatures this becomes
\begin{equation}
\lambda_m \to T^\frac{1-d}{2}\,.
\end{equation}

\subsubsection{Asymptotic form of the eigenvalue distribution}
For completeness, we give the asymptotic form of the eigenvalue distribution for $T \gg T_c$. 
In the vector model, we have
\begin{equation}
\rho(\lambda) = \frac{A_d}{2} \left(\frac{T}{T_c}\right)^{d-1}
\begin{cases}
\pi(\lambda_m - |\lambda|)&\text{for } d = 3\,,\\
\frac{1}{2}\left(\lambda_m^2(3-2\log\lambda_m)-\lambda^2(3-2\log|\lambda|)\right)&\text{for } d = 4\,,\\
\zeta(d-3) (\lambda_m^2 - \lambda^2)&\text{for } d \geq 5\,,
\end{cases}
\end{equation}
yielding the asymptotic Fourier coefficients
\begin{multline}
	\label{eq:asymfourier}
\rho_k = 2 A_d \left(\frac{T}{T_c}\right)^{d-1} \frac{1}{k^2}\\
\times\begin{cases}
\pi \sin^2\left(\frac{k\lambda_m}{2}\right)&\text{for } d = 3\,,\\
\frac{1}{k}(\text{Si}(k\lambda_m)-\log\lambda_m \sin(k\lambda_m))+\lambda_m\cos(k\lambda_m)(\log\lambda_m - 1)&\text{for } d = 4\,,\\
\zeta(d-3) \left(\frac{\sin(k\lambda_m)}{k} - \lambda_m\cos(k\lambda_m)\right)&\text{for } d \geq 5\,,
\end{cases}
\end{multline}
where $\text{Si}(z) \equiv \int_{0}^{z} dt\,t^{-1}\sin t$. 
This implies a $k^{-2}$-falloff for large $k \gg \lambda_m^{-1}$ in all dimensions.

In the adjoint model, the distribution approaches
\begin{equation}
\rho(\lambda) = \frac{2}{\pi \lambda_m^2}\sqrt{\lambda_m^2 - \lambda^2}\,,
\end{equation}
which leads to the Fourier moments
\begin{equation}
\rho_k = \frac{2}{k\lambda_m} J_1(k \lambda_m)\,,
\end{equation}
with the first Bessel function of the first kind $J_1$. Here, the large $k \gg \lambda_m^{-1}$ falloff is $k^{-3/2}$.

\section{The thermal Green's function}  
\label{sec:thermalgreen}  

The scalar Green's function for a particular eigenvalue distribution $\lambda_i$ can be read off 
directly from Eq.~\eqref{eq:scal_ac}. 
The inversion of the kinetic operator is most readily implemented by decomposing it in 
terms of its eigenvalues and eigenvectors. In the vector model, this yields
\begin{align}
G_{AB}(0,y) =&
\beta^{-1}\sum_{i,n,\ell,M} \frac{Y_{\ell,M}(0) Y_{\ell,M}^*(y) \Psi^{i}_A (\Psi^{i}_B)^\dagger }
{\beta^{-2}\left(2\pi n + \lambda_i\right)^2 + (\ell+(d-2)/2)^2}\nonumber\\
=& (2\pi\beta)^{-1}\frac{\Gamma(\sigma)}{\pi^{\sigma}}
\sum_{i,n,\ell} \frac{(\ell+\sigma) C_\ell^{(\sigma)}(\cos\theta) \Psi^{i}_A (\Psi^{i}_B)^\dagger}
{\beta^{-2}\left(2\pi n + \lambda_i\right)^2 + (\ell+\sigma)^2}\,,
\label{eq:gf}
\end{align}
where $0$ and $y$ refer to points on $S^{d-1}$ separated by a polar angle $\theta$.
Here, $i$ runs from $1$ to $N$, $n$ from $-\infty$ to $\infty$, and $\ell$ from $0$ to $\infty$. 
$M$ denotes $d-2$ magnetic indices that obey $\ell \geq |m_{d-2}| \geq ... \geq |m_1|$.
$Y_{\ell,M}$ are the $d$-dimensional spherical harmonics. We have used their 
orthogonality and the fact that one insertion point is at 0 to set $M=0$ to obtain 
the second line of Eq.~\eqref{eq:gf}. They are eigenfunctions 
of the spherical part of the Laplacian with eigenvalues $-\ell(\ell+d-2)$. We have introduced 
$\sigma=(d-2)/2$ for notational simplicity and $C_\ell^{(\sigma)}$ are the Gegenbauer 
polynomials which reduce to the regular Legendre polynomials for $d=3$. 
$\Psi^{i}_A$ is the $i$-th eigenvector of the holonomy matrix $U$ at the saddle point, 
with eigenvalue $e^{i\lambda_i}$. We have also included color indices $A$, $B$ for clarity. 
The eigenvectors are normalized according to
\begin{equation}
\label{eq:evec_norm}
\Psi^{i}_A (\Psi^{j}_A)^\dagger = \delta^{ij}\,,\hspace{3em}
\Psi^{i}_A (\Psi^{i}_B)^\dagger = \delta_{AB}\,.
\end{equation}
The sum over $\ell$ can be performed explicitly.
To this end, we first rewrite the sum over $n$ as\footnote{Here, we have made use 
of the identity \begin{equation}\sum_{n=-\infty}^\infty \frac{E}{\left(2\pi n + \lambda\right)^2 
+ E^2} = \frac{1}{2}\sum_{n=-\infty}^\infty e^{-|n|E} \cos{n \lambda}\,.\end{equation}}
\begin{equation}\label{eq:gffunexp}
G_{AB}(0,y) =
\frac{\Gamma(\sigma)}{4\pi^{\sigma+1}}
\sum_j\Psi^{j}_A (\Psi^{j}_B)^\dagger
\sum_{n,\ell} C_\ell^{(\sigma)}(\cos\theta)\, e^{-\beta|n|(\ell+\sigma)} \cos{n \lambda_j}\,,
\end{equation}
and then use the generating functional for Gegenbauer polynomials,
\begin{equation}
\label{eq:leg_gen}
\frac{1}{\left(1-2 t x+t^2\right)^\sigma} = \sum_{\ell=0}^{\infty} t^\ell C_\ell^{(\sigma)}(x)\,,
\end{equation}
to sum over $\ell$ and obtain
\begin{equation}
\label{eq:gffun}
G_{AB}(0,y) =
\frac{1}{2}\frac{\Gamma(\sigma)\ }{(2\pi)^{\sigma+1}}
\sum_j\Psi^{j}_A (\Psi^{j}_B)^\dagger
\sum_{n} \frac{\cos{n \lambda_j}}{\left(\cosh n\beta - \cos\theta\right)^\sigma}\,.
\end{equation}
As one might have anticipated, for $\lambda_j = 0$ this agrees with the expression 
obtained by considering the Green's function in the universal covering space $R \times S^{d-1}$ 
and subsequently implementing periodic boundary conditions for the thermal direction by 
summing over all images.

In the adjoint case, the corresponding Green's function reads 
\begin{equation}
\label{eq:gfadj}
G_{AB,CD}(0,y) = \frac{1}{2}\frac{\Gamma(\sigma)}{(2\pi)^{\sigma+1}}
\sum_{j\neq k} \Psi^{j}_A (\Psi^{k}_B)^\dagger \Psi^{k}_C (\Psi^{j}_D)^\dagger \sum_n \frac{\cos{n (\lambda_j-\lambda_k)}}{\left(\cosh n\beta - \cos\theta\right)^\sigma}\,.
\end{equation}

\section{Two-point functions of local operators A: The vector model}
\label{vectormodel_2pt}

Let us now consider expectation values of the form 
$\exv{\text{Tr}\, |\phi(0)|^2\,\text{Tr}\,|\phi(y)|^2}$.
As usual, by Wick's theorem, expectation values are obtained via operator contractions
and in this case normal ordering and symmetries leave only one 
nontrivial contraction,
\begin{equation}
\big\langle\text{Tr}\, |\phi(0)|^2\,\text{Tr}\,|\phi(y)|^2\big\rangle = 
\contraction{\,\,\langle\text{Tr}\,}{\phi}{^\dagger(0)\phi(0)\text{Tr}\,\phi^\dagger(y)}{\phi}
\contraction[2ex]{\,\,\langle\text{Tr}\,\phi^\dagger(0)}{\phi}{(0)\text{Tr}\,}{\phi}
\left\langle\text{Tr}\, \phi^\dagger(0) \phi(0)\text{Tr}\, \phi^\dagger(y) \phi(y)\right\rangle\,
=\, 
\frac{1}{N}G_{AB}(0,y)G_{BA}(0,y)\,.
\label{2pt_vector}
\end{equation}
In order to facilitate making contact with bulk duals, we have chosen to normalize the fields so that 
their vacuum two point functions are independent of $N$. This explains the factor of $1/N$ 
on the right hand side of the second equality in Eq.~\eqref{2pt_vector}. 
Inserting Eq.~\eqref{eq:gffun} and 
employing Eq.~\eqref{eq:evec_norm}, we obtain
\begin{align}
\big\langle\text{Tr}\, |\phi(0)|^2\,\text{Tr}\,|\phi(y)|^2\big\rangle =&\,
\frac{\Gamma^2(\sigma)}{4(2\pi)^{d}N}
\sum_i\,
\left(\sum_{n} \frac{\cos{n \lambda_i}}{\left(\cosh n\beta 
- \cos\theta\right)^\sigma}\right)^2\nonumber\\
\label{eq:4ptcorrfour}
=&\,\frac{\Gamma^2(\sigma)}{2^{d+2}\pi^{d}}
\sum_{n,m = -\infty}^\infty 
\frac{\rho_{|n-m|}}{\left[(\cosh n\beta-\cos\theta)(\cosh m\beta-\cos\theta)\right]^\sigma} \,,
\end{align}
where we have used Eq.~\eqref{eq:rho_k} on the second line. The two-point function can now be 
approximated by inserting the saddle point values of $\rho_n$. 


\subsection{Low temperature phase: $T \leq T_c$}
Below the phase transition, $\rho_n$ is given by 
\begin{equation}
\rho_{n}=\begin{cases} \ \ 1&\text{ for } n=0\,,\\
\frac{N_f}{N} z_S^d(x^{n}) &\text{ for } n> 0\,.
\end{cases}  
\end{equation}
Inserting this into the general 
expression Eq.~\eqref{eq:4ptcorrfour} for the two-point function, gives
\begin{align}
\big\langle\text{Tr}\, |\phi(0)|^2\,\text{Tr}\,|\phi(y)|^2\big\rangle &
= \frac{\Gamma^2(\sigma)}{2^{d+2}\pi^{d}}\Bigg(
\sum_{n = -\infty}^\infty (\cosh n\beta-\cos\theta)^{2-d}\nonumber\\
&\hspace{6em}+\frac{N_f}{N}\sum_{n \neq m} 
\frac{z_S^d(x^{|n-m|})}{\left[(\cosh n\beta-\cos\theta)(\cosh m\beta-\cos\theta)\right]^\sigma}
\Bigg)\nonumber\\
&= {\cal K}^{d+1}_\text{AdS}(\beta,\theta) + {\cal O}(1/N)\,,
\label{leading_2pt}
\end{align}
where ${\cal K}^d_\text{AdS}(\beta,\theta)$ is the boundary-to-boundary correlator in 
$d$-dimensional thermal AdS space.
We will explore this connection to thermal AdS correlators in more detail in 
section~\ref{sec:discussion}. 

The next step is consider the behavior of the the two-point function in different distance regimes.

\subsubsection{Short distance $\theta\ll\beta$}

In the $\theta\ll\beta$ limit, the dominant terms are those singular in $\theta$. 
They arise exclusively from terms with $m = 0$ and/or $n = 0$ in Eq.~\eqref{eq:4ptcorrfour}. 
In the short distance limit the two-point function is thus given by  
\begin{align}
\exv{\text{Tr}\, |\phi(0)|^2\,\text{Tr}\,|\phi(y)|^2} =&\frac{\Gamma^2(\sigma)}{2^{d+2}\pi^{d}}
\Bigg(\frac{1}{\left(1-\cos\theta\right)^{2\sigma}}
+\frac{4}{\left(1-\cos\theta\right)^\sigma}
\sum_{n = 1}^\infty \frac{\rho_{n}}{\left(\cosh n\beta-\cos\theta\right)^\sigma}\Bigg) \nonumber\\
\label{eq:4ptcorrshort}
\approx&\,
\frac{\Gamma^2(\sigma)}{2^{d+2}\pi^{d}}\left( \frac{1}{\left(1-\cos\theta\right)^{2\sigma}}
+\frac{N_f}{N}\frac{2^{\sigma+2}}{\left(1-\cos\theta\right)^\sigma}
\sum_{n = 1}^\infty z_S^{2(d-1)}(x^n)\right).
\end{align}
The leading contribution at any temperature comes from the vacuum piece and reads
\begin{equation}\label{eq:shortlimit}
\exv{\text{Tr}\, |\phi(0)|^2\,\text{Tr}\,|\phi(y)|^2}  \approx
\frac{\Gamma^2(\sigma)}{16\pi^{d}} \frac{1}{\theta^{2(d-2)}}\,.
\end{equation}

\subsubsection{Long distance $\theta\gg\beta$}

Since $\theta\leq\pi$ it follows that long distance compared to $\beta$ implies that 
$\beta\ll 1$, {\it i.e.} that we are at relatively high temperature
even if we are still below $T_c$. 
Due to the suppression of the higher Fourier coefficients of $\rho$, we can 
expand the denominator of Eq.~\eqref{eq:4ptcorrfour} in $(n-m)\beta$ and arrive, 
after shifting the summation index $m$, at
\begin{align}\label{eq:4ptcorrlong}
\exv{\text{Tr}\, |\phi(0)|^2\,\text{Tr}\,|\phi(y)|^2}
&\approx \frac{\Gamma^2(\sigma)}{2^{d+2}\pi^{d}}
\left(\sum_{p = -\infty}^{\infty}\rho_{|p|}\right)
\sum_{n = -\infty}^\infty (\cosh n\beta-\cos\theta)^{2-d}\nonumber\\
&\approx \left(1+4\,\zeta(d{-}1)\frac{N_f}{N}\beta^{1-d}\right){\cal K}^{d+1}_\text{AdS}(\beta,\theta)
\,.
\end{align}
This reveals an interesting fact. Even when the ${\cal O}(1/N)$ terms in 
Eq.~\eqref{leading_2pt} begin to contribute, they do so in a controlled manner, 
as long as we are below the phase transition. Indeed, they only provide a temperature 
dependent renormalization of the leading order result.


\subsection{High temperature phase: $T > T_c$}

In the high temperature phase, the behavior of the correlators is more subtle 
than below the transition. The reason for this is the emergence of an additional length scale 
in the sum in Eq.~\eqref{eq:4ptcorrfour} once a hierarchy between $T$ and $T_c$ opens up, 
{\it i.e.} for $\lambda_m \ll 1$. 

We can easily infer the emergence of the new length scale from the Fourier form of the 
eigenvalue distribution in Eq.~\eqref{eq:asymfourier}. For $n \ll \lambda_m^{-1}$ it is 
independent of $n$ to leading order, while for $n \gg \lambda_m^{-1}$ it is oscillatory with 
an envelope that scales as $n^{-2}$. More precisely, using the high temperature expression for 
$\rho_n$, we have
\begin{equation}
\rho_n \to 1\text{ for } n\lambda_m \ll 1\,.
\end{equation}
Thus, whenever the correlator is dominated by $n \ll \lambda_m^{-1}$, it corresponds 
to that of an unconstrained, free theory. On the other hand, for $n \gg \lambda_m^{-1}$, 
the full expression for the distribution function is relevant and when this regime dominates 
a correlation function, the singlet constraint will be important and we will find behavior 
similar to the low temperature phase.

To find the crossover scale, we should not compare $n$ itself to the angle $\theta$, 
but instead the combination $n\beta$, since this is what enters in the denominators in 
Eq.~\eqref{eq:4ptcorrfour}. This yields a crossover scale $\theta_c \approx \beta\lambda_m^{-1}$. 
Inserting the asymptotic expression for the maximal angle $\lambda_m$, the following 
picture emerges. In $d=3$, $\lambda_m \sim \beta_c/\beta$ and the crossover scale is 
independent of $\beta$ and at a rather small angle,  $\theta_c \sim 1/\sqrt{N}$.
In higher dimensions, on the other hand, the crossover scale grows with temperature. 
At sufficiently high $T$, it surpasses the size of the sphere, and no regime 
$\theta > \theta_c$ exists. Therefore, in higher dimensions, the large distance 
correlator above the phase transition differs significantly from its low temperature 
counterpart. In particular, at large enough temperature it cannot be identified with the boundary-to-boundary correlator 
in thermal AdS.

\subsubsection{Below the crossover scale: $\theta\ll\theta_c$}

The crossover in angular behavior occurs at $\theta_c \approx \beta\lambda_m^{-1}$, which 
depends on temperature as follows in different dimensions,
\begin{equation}\label{eq:theta_c}
\theta_c \sim \beta_c \times
\begin{cases}
1&\text{for }d=3\,,\\
\log^{1/3}\frac{\beta_c}{\beta}&\text{for }d=4\,,\\
\left(\frac{\beta_c}{\beta}\right)^{\frac{d-4}{3}} &\text{for }d\geq 5\,. 
\end{cases}
\end{equation}
Taking a look at the continuum version of Eq.~\eqref{eq:gffunexp},
\begin{equation}
\frac{\Gamma(\sigma)}{4\pi^{\sigma+1}}
\int d\lambda\,\rho(\lambda)
\sum_{n,l} C_l^{(\sigma)}(y)\, e^{-\beta|n|(l+\sigma)} \cos{n \lambda}\,,
\end{equation}
we observe that relevant contributions arise only from $n \lesssim 1/\beta E_l$, with 
$E_l=l+\sigma$, since larger $n$ are exponentially suppressed. 
In the relevant range we have 
\begin{equation}
n\lambda \leq n\lambda_m \lesssim \frac{1}{\beta_c E_l} \sim \frac{\theta}{\theta_c} \ll 1\,,
\end{equation}
which allows us to approximate $\cos n\lambda \approx 1$ and thus $\rho_n \approx 1$. 
In this regime, the distribution is therefore $\delta$-like and we obtain
\begin{equation}
\label{eq:id_fun}
\exv{\text{Tr}\, |\phi(0)|^2\,\text{Tr}\,|\phi(y)|^2} 
=\,\frac{\Gamma^2(\sigma)}{2^{d+2}\pi^{d}}\left(\sum_n \left[\cosh n\beta 
- \cos\theta\right]^{-\sigma}\right)^2\,.
\end{equation}

\paragraph{Subthermal distances, $\theta \ll \beta$:}

In the short distance regime, we again focus on the contributions to Eq.~\eqref{eq:id_fun} 
that are singular for $\theta \to 0$,
\begin{equation}
\exv{\text{Tr}\, |\phi(0)|^2\,\text{Tr}\,|\phi(y)|^2}
\approx\, 
\frac{\Gamma^2(\sigma)}{2^{d+2}\pi^{d}}\Bigg(\frac{1}{\left(1-\cos\theta\right)^{2\sigma}}+\frac{4}{\left(1-\cos\theta\right)^\sigma}\sum_{n=1}^{\infty} \frac{1}{\left(\cosh n\beta - 1\right)^\sigma}\Bigg)\,.
\end{equation}
Again, the leading contribution is given by Eq. \eqref{eq:shortlimit} and coincides with the low temperature phase short distance behavior.

\paragraph{Intermediate distances, $\beta \ll \theta \ll \theta_c$:}

To understand the behavior in this regime, we leave $S^{d-1}$ for a moment and consider flat space instead. There, the equivalent of Eq.~\eqref{eq:id_fun} can be written in a momentum representation as
\begin{align}
\exv{\text{Tr}\, |\phi(0)|^2\,\text{Tr}\,|\phi(y)|^2} &\sim \left(\int d^{d-1}k \,e^{iky}\sum_{n=-\infty}^{\infty} e^{-\beta n k} \right)^2\nonumber\\
&\sim \left(\int d^{d-1}k \,(1+2n_B\left(k\right)) e^{iky}\right)^2\,,
\end{align}
where we have approximated $E_k \sim k$ and evaluated the sum over $n$ to arrive at the second line.
Since we are at scales much larger than the thermal wavelength, $k \beta \ll 1$, we can expand the Bose distribution to yield
\begin{align}
\exv{\text{Tr}\, |\phi(0)|^2\,\text{Tr}\,|\phi(y)|^2}
&\sim \left(\int dk\, k^{d-2}\left(1+ \frac{2}{\beta k} \right) e^{iky}\right)^2\nonumber\\
&\sim \left(\frac{1}{|y|^{d-1}} + \frac{1}{\beta |y|^{d-2}} \right)^2 \sim \frac{1}{\beta^2 |y|^{2(d-2)}}\,,
\end{align}
where the final relation holds for $\beta \ll |y|$. The result is nothing but the vacuum 
correlator in $d-1$ dimensions. We thus observe an effective dimensional reduction 
of the correlator in the intermediate regime. Back on the sphere, we can perform the sum in Eq. \eqref{eq:id_fun}. If we further take the small angle approximation, keeping in mind that $1\gg\theta\gg\beta$, then the leading contribution simply reads  
\begin{equation}
\exv{\text{Tr}\, |\phi(0)|^2\,\text{Tr}\,|\phi(y)|^2}  \sim 
\begin{cases}
\frac{1}{\beta^2}\log^2\theta &\text{for } d = 3\,,\\
\frac{1}{\beta^2}\theta^{2(3-d)} &\text{for } d > 3\,.
\end{cases}
\end{equation}

\subsubsection{Long distances: $\theta>\theta_c$}

As discussed above, this regime exists only as long as $\theta_c$ remains small. This holds
at all temperatures in $d=3$ but in higher dimensions the temperature range is restricted.
It is straightforward to check from Eq.~\eqref{eq:theta_c}, that the allowed range is 
$\beta_c\geq\beta\gg\beta_c '$, where
\begin{equation}
	\beta_c ' \sim \begin{cases}
0 &\text{for } d = 3\,,\\
\beta_c \,e^{-1/\beta_c^3}\sim N^{-\frac{1}{3}} e^{-N} &\text{for } d = 4\,,\\
\beta_c^{\frac{d-1}{d-4}} \sim N^{-\frac{1}{d-4}} &\text{for } d \geq 5\,.
	\end{cases}
\end{equation}
For $d=3$ and $d=4$ this regime is available at basically all temperatures above the 
phase transition, whereas for $d\geq5$, the higher the dimension the narrower the 
temperature range for which the critical angle is small. We note, however, that in any 
given dimension $d$ the temperature range, where we can observe the long distance 
behavior at $\theta>\theta_c$, gets wider as we take $N$ larger.

Within this temperature interval, the long distance correlator takes a surprisingly
simple form. We can apply the same expansion that lead to the first line of
Eq.~\eqref{eq:4ptcorrlong}, and obtain an expression of the same form,
\begin{equation}
\exv{\text{Tr}\, |\phi(0)|^2\,\text{Tr}\,|\phi(y)|^2}
\approx \left(\sum_{p = -\infty}^{\infty}\rho_{|p|}\right){\cal K}^{d+1}_\text{AdS}(\beta,\theta)\,. \label{innerregion}
\end{equation}
The multiplicative factor in front of the AdS propagator depends on temperature but not
on the angle $\theta$, so the leading behavior at long distances once again corresponds to 
propagation in thermal AdS, up to a temperature dependent normalization factor. 

Let us summarize the high temperature behavior above the phase transition. We find that 
in the limit of small operator separation the correlation function has rather simple angular 
dependence, that corresponds to propagation in thermal AdS space of one higher dimension.
When $\theta>\beta$ the separation between the operator insertions is larger than the thermal 
length and the correlator behaves in a qualitatively different manner, reflecting an effective
dimensional reduction. However, when $d$ is low and the temperature is not too high, we
observe an unexpected departure from this behavior at the largest 
available angular separations. Surprisingly, the higher dimensional thermal AdS propagator
is recovered up to an overall temperature dependent normalization in this regime. 

It is not immediately clear how to interpret these results from the point of view of 
gauge/gravity duality, although the behavior of the two-point correlation function at low 
temperatures and at short distances is suggestive of an emerging AdS geometry. 
It is then tempting to attribute the deviation from AdS behavior at temperatures above the 
phase transition to an emerging geometry of an extended object, 
possibly a very large black hole (much larger than AdS scale), but the large angle behavior, 
where we recover a form of thermal AdS propagation, is rather mysterious from that
point of view. 

\section{Two-point functions of local operators B: The adjoint model}
\label{sec: adjointmodel_2pt}

We again limit our attention to the only non-trivial Wick contraction and normalize the 
field as before,
\begin{equation}
\exv{\text{Tr}\, |\phi(0)|^2\,\text{Tr}\,|\phi(y)|^2}=
\,\frac{\Gamma^2(\sigma)}{2^{d+2}N^2\pi^{d}}\,\sum_{i,j}
\left(\sum_{n} \frac{\cos{(n(\lambda_i-\lambda_j))}}{\left(\cosh n\beta - \cos\theta\right)^\sigma}\right)^2\,.
\end{equation}
Analogous to the vector model, we can write this as
\begin{equation}\label{eq:2pointadj}
\exv{\text{Tr}\, |\phi(0)|^2\,\text{Tr}\,|\phi(y)|^2} =
\frac{\Gamma^2(\sigma)}{2^{d+2}\pi^{d}}
\sum_{n,m=-\infty}^\infty \frac{\rho_{|m-n|}^2}{\left[(\cosh m\beta-\cos\theta)(\cosh n\beta-\cos\theta)\right]^\sigma}\,,
\end{equation}
where we have taken the eigenvalue distribution to be symmetric around $\lambda = 0$, 
without loss of generality.
Note the quadratic dependence on the $\rho_n$. This reflects the $U(1)$ (or, for finite $N$ 
the $\mathbb{Z}_N$ center) symmetry of the theory\footnote{As mentioned in 
Section~\ref{sec:saddle}, integration over the center gives rise to an additional overall factor of $2\pi$. 
We have normalized our correlation function in such a way that the vacuum contribution is  
given by the boundary-to-boundary correlator in $d{+}1$-dimensional thermal AdS space.}. 

\subsection{Low temperature phase: $T \leq T_c$}

Below the transition, due to the vanishing of all off-diagonal contributions, the two-point function 
simplifies to\footnote{Subdominant pieces in $1/N$ arise from taking fluctuations around 
the saddle into account. Due to the Gaussian nature of the partition function 
Eq.~\eqref{eq:act_adj} this is straightforward. We obtain
\begin{equation}
\exv{\rho_n^2} = \frac{1}{2 N^2 (1 - z^d_S(x^n))}\,,
\end{equation}
and therefore for the subleading piece
\begin{equation}
\frac{1}{2}\sum_{n,m=-\infty}^\infty (1 - z^d_S(x^{|m-n|}))^{-1}
\left[(\cosh m\beta-\cos\theta)(\cosh n\beta-\cos\theta)\right]^{-\sigma}\,.
\end{equation}
}
\begin{equation}
\exv{\text{Tr}\, |\phi(0)|^2\,\text{Tr}\,|\phi(y)|^2}
=\frac{\Gamma^2(\sigma)}{2^{d+2}\pi^{d}}\sum_{n=-\infty}^\infty 
\left(\cosh n\beta-\cos\theta\right)^{2-d} = {\cal K}^{d+1}_\text{AdS}(\beta,\theta)\,.
\end{equation}
Thus, once more, this corresponds to the boundary-to-boundary correlator in thermal AdS.
At very low temperatures, $\cosh n\beta \sim \frac12 e^{n\beta}$ and the finite temperature 
part of the above expression is exponentially suppressed,
\begin{equation}
\sum_{n=1}^{\infty} \left(\cosh n\beta-\cos\theta\right)^{2-d}
\approx 2^{d-2}  e^{-(d-2)\beta}\left(1+2 (d-2) e^{-\beta}\cos\theta\right)\,.
\end{equation}

\subsection{High temperature phase: $T > T_c$}

Around the phase transition, the exact form of the eigenvalue distribution is 
somewhat complicated but it simplifies significantly in the very 
high temperature regime.
Similar to the vector model considered in Section~\ref{vectormodel_2pt}, there is a 
crossover scale $\theta_c$ in the adjoint model at which the correlator crosses over 
from that of an unconstrained theory to that of a theory with a singlet condition.

\subsubsection{Below the crossover scale: $\theta \ll \theta_c$}
One again finds that $\theta_c = \beta\lambda_m^{-1}$, but the
expression for $\lambda_m$ differs from the vector model. At sufficiently high temperature,
we find 
\begin{equation}
\lambda_m \approx \beta^{\frac{d-1}{2}}\,,
\end{equation}
and thus
\begin{equation}
\theta_c \approx \beta^{\frac{3-d}{2}}\,.
\end{equation}
As before, we can approximate the eigenvalue distribution at angles below the crossover scale
by a $\delta$-function, {\it i.e.} $\rho_n = 1$ for all $n$, and obtain
\begin{equation}
\sum_{mn} \frac{\rho_{|n-m|}^2}{\left[(\cosh m\beta-\cos\theta)(\cosh n\beta-\cos\theta)\right]^{\sigma}} 
\approx \left(\sum_{n} \left(\cosh n\beta-\cos\theta\right)^{-\sigma}\right)^2\,.
\end{equation}
As in the vector model, there are two cases to consider.

\paragraph{Subthermal distance, $\theta \ll \beta$:}

The leading contributions are again given by the singular terms. Correspondingly, we obtain
\begin{equation}
\exv{\text{Tr}\, |\phi(0)|^2\,\text{Tr}\,|\phi(y)|^2}  \approx
\frac{\Gamma^2(\sigma)}{2^{d+2}\pi^{d}} \left(\frac{1}{\left(1-\cos\theta\right)^{2\sigma}}+ 
\frac{4}{\left(1-\cos\theta\right)^{\sigma}}\sum_{n=1}^\infty 
\frac{1}{\left(\cosh n\beta -1\right)^{\sigma}}\right)
\end{equation}
Thus the leading behavior at sufficiently small angles, $\theta \ll 1$, is exactly the same as in the
vector model,
\begin{equation}
\exv{\text{Tr}\, |\phi(0)|^2\,\text{Tr}\,|\phi(y)|^2}  \approx
\frac{\Gamma^2(\sigma)}{16\pi^{d}} \frac{1}{\theta^{2(d-2)}}\,.
\end{equation}

\paragraph{Intermediate distance, $\beta \ll \theta \ll \theta_c$:}

For the same reasons as in the vector model, intermediate distances 
imply dimensional reduction and we find that
\begin{equation}
\exv{\text{Tr}\, |\phi(0)|^2\,\text{Tr}\,|\phi(y)|^2}  \sim 
\begin{cases}
\frac{1}{\beta^2}\log^2\theta &\text{for } d = 3\,,\\
\frac{1}{\beta^2}\theta^{2(3-d)} &\text{for } d > 3\,.
\end{cases}
\end{equation}

\subsubsection{Long distance $\theta > \theta_c$}

In contrast to the vector model, the critical temperature is now $\beta_c\sim\mathcal{O}(1)$, 
implying that close to the phase transition the crossover angle is $\theta_c \sim\mathcal{O}(1)$ 
in any dimension. For $d>3$ the crossover angle grows with temperature above the phase
transition and becomes larger than $\pi$ at rather small temperature, so no simple temperature expansion can be taken.
For $d=3$, on the other hand, the crossover angle remains at $\mathcal{O}(1)$ at high 
temperatures and in this special case, an expansion of Eq.~\eqref{eq:2pointadj} in powers 
of $\beta(m-n)$ leads to 
\begin{equation}
\exv{\text{Tr}\, |\phi(0)|^2\,\text{Tr}\,|\phi(y)|^2}
\approx \left(\sum_{p = -\infty}^{\infty}\rho^2_{|p|}\right){\cal K}^{4}_\text{AdS}(\beta,\theta)\,,
\end{equation}
for $\theta>\theta_c$. This result is qualitatively similar to Eq.~\eqref{innerregion} and 
recovering higher-dimensional thermal AdS behavior at large angles is equally mysterious 
here.

\section{Summary and discussion}\label{sec:discussion}

\begin{figure}[t]
	\centering{
		\includegraphics[width=.49\textwidth]{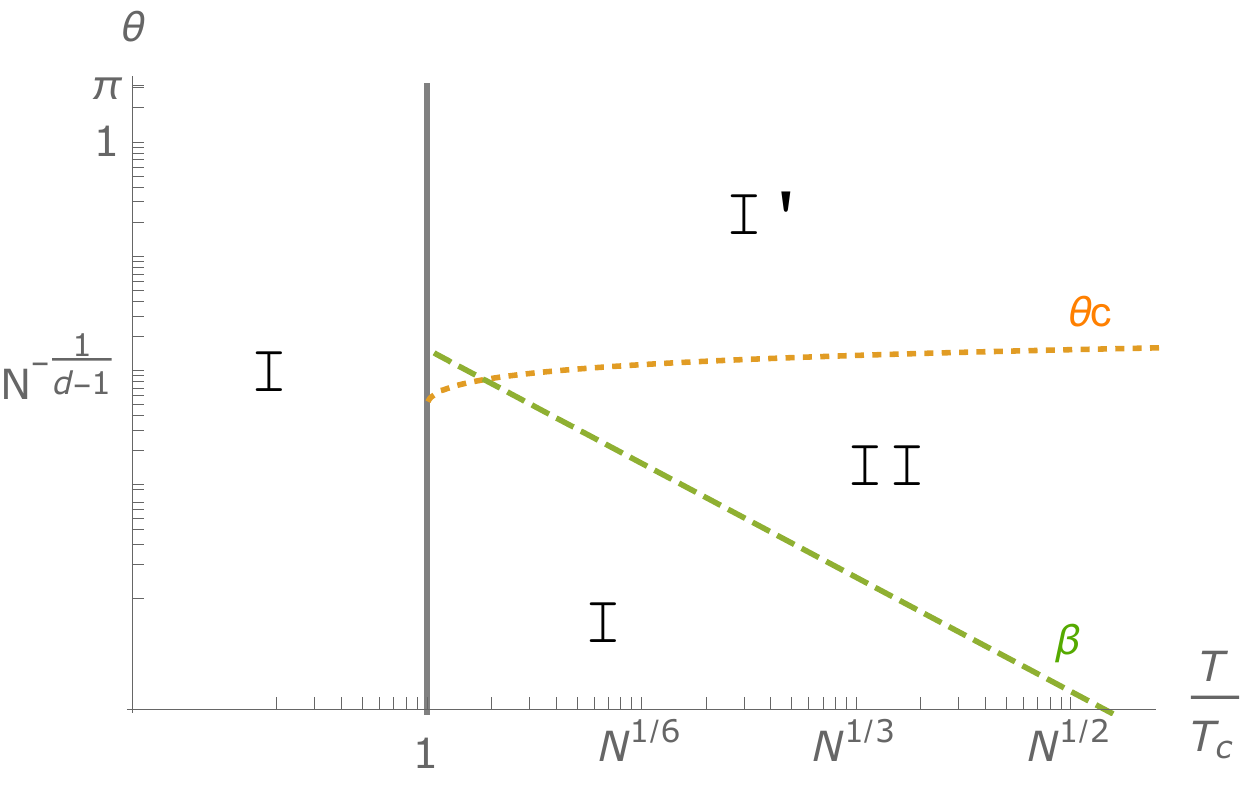}\hfill
	    \includegraphics[width=.49\textwidth]{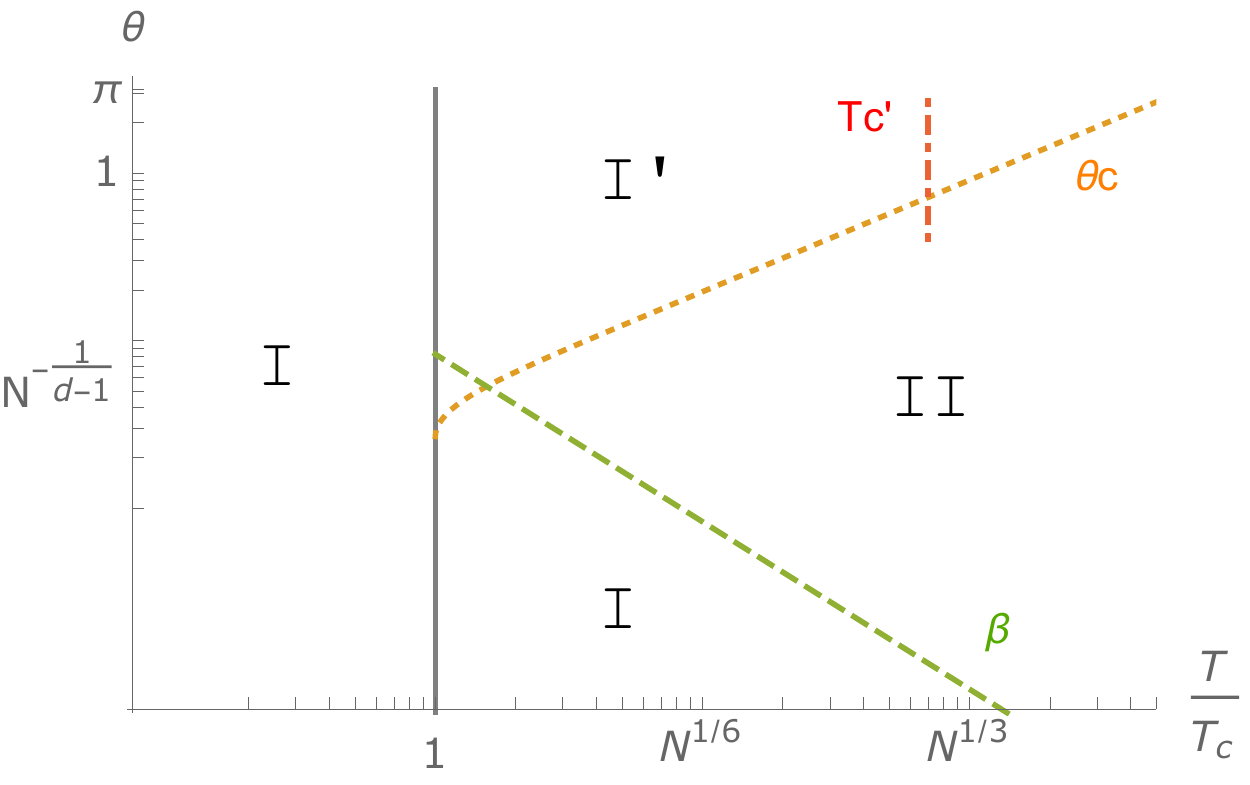}
		\caption{\label{fig:phasediagram}Vector model phase diagram for $d=3,4$ (left) and $d\geq5$ (right). In regions~I, ~I' and~II the correlation functions are given by ${\cal K}^{d+1}_\text{AdS}(\beta,\theta)$, $f_\beta{\cal K}^{d+1}_\text{AdS}(\beta,\theta)$, and $\frac{F_d (\theta)}{\beta^2}$, respectively.  Unlabeled regions correspond to crossover regions without a simple approximation for the correlators.
		}
	}
\end{figure}

\begin{figure}[t]
	\centering{
		\includegraphics[width=.49\textwidth]{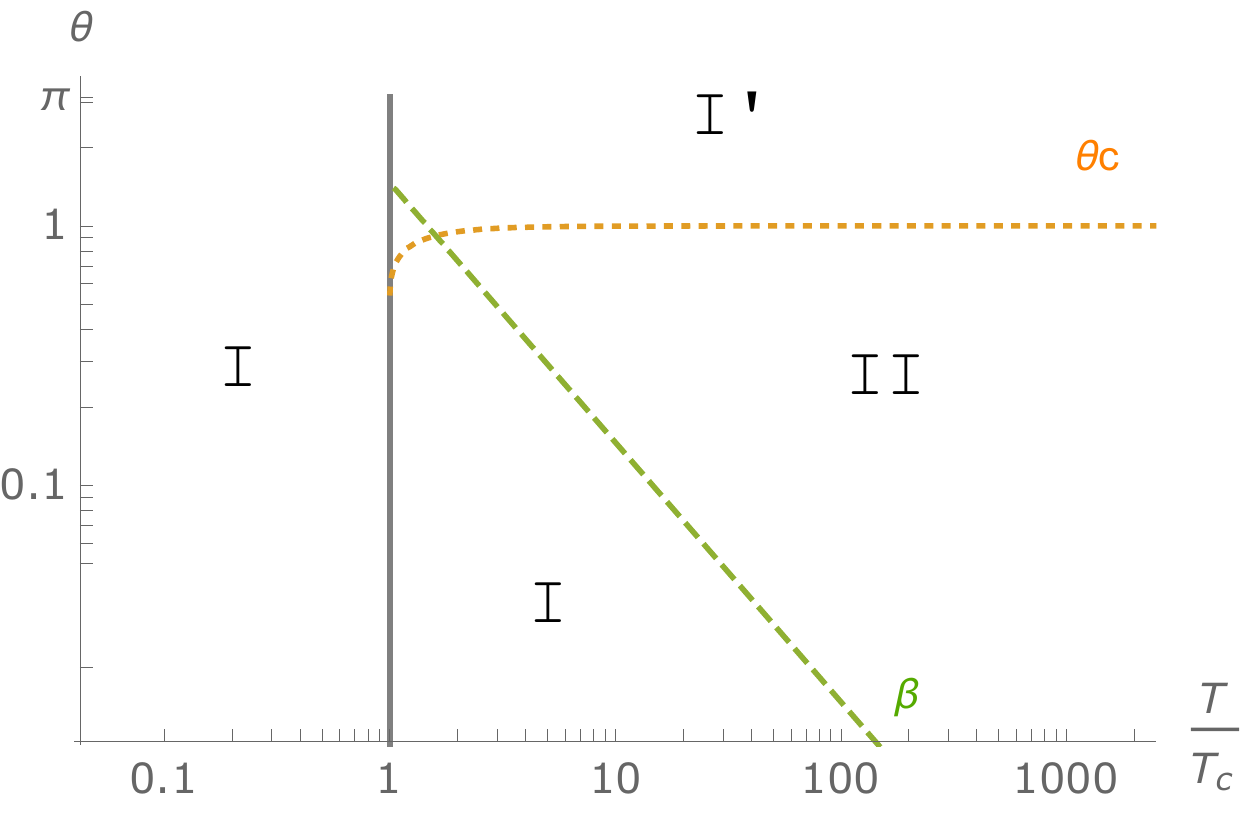}\hfill
		\includegraphics[width=.49\textwidth]{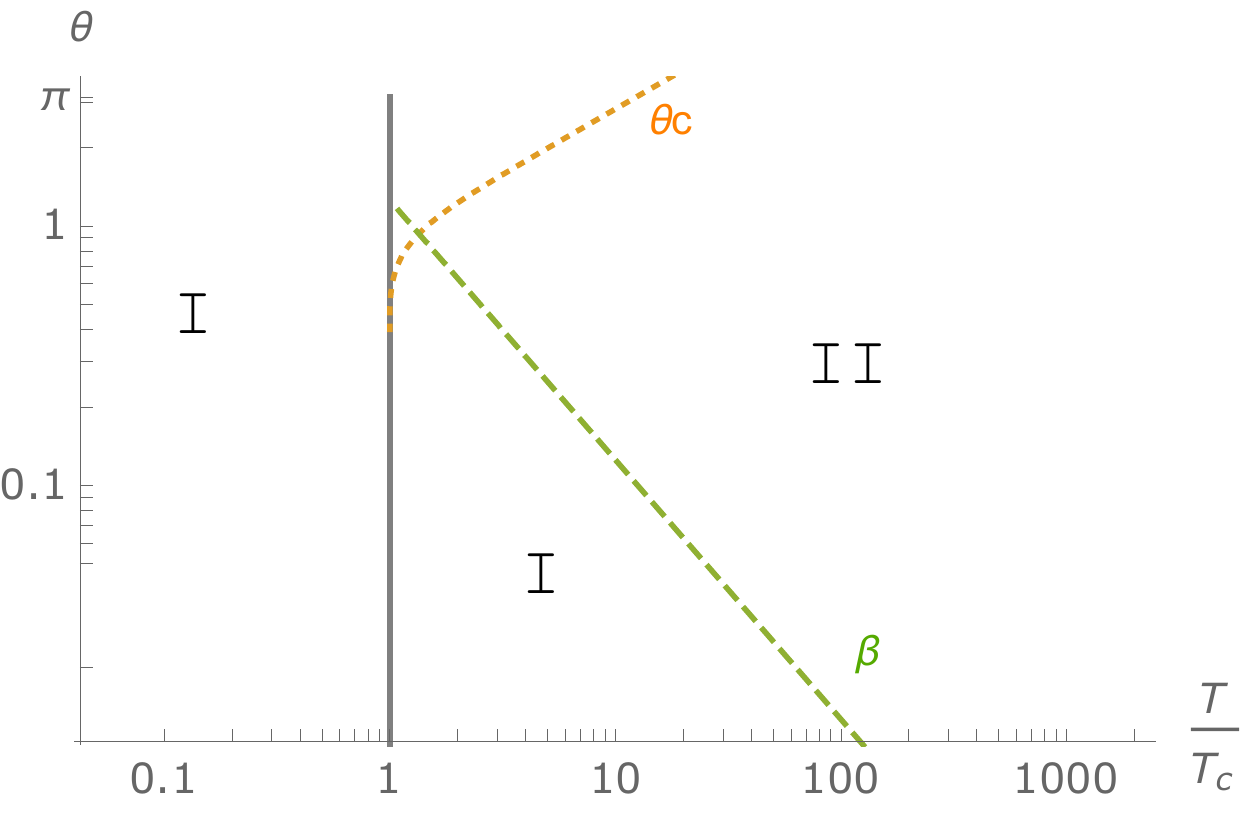}
		\caption{\label{fig:phasediagramAdj}Adjoint model phase diagram for $d=3$ (left) and $d\geq4$ (right). In regions~I,~I' and II the correlation functions are given by ${\cal K}^{d+1}_\text{AdS}(\beta,\theta)$, $f_\beta{\cal K}^{d+1}_\text{AdS}(\beta,\theta)$, and $\frac{F_d (\theta)}{\beta^2}$, respectively.  Unlabeled regions correspond to crossover regions without a simple approximation for the correlators.
		}
	}
\end{figure}

\begin{figure}[t]
	\centering{
		\includegraphics[width=.49\textwidth]{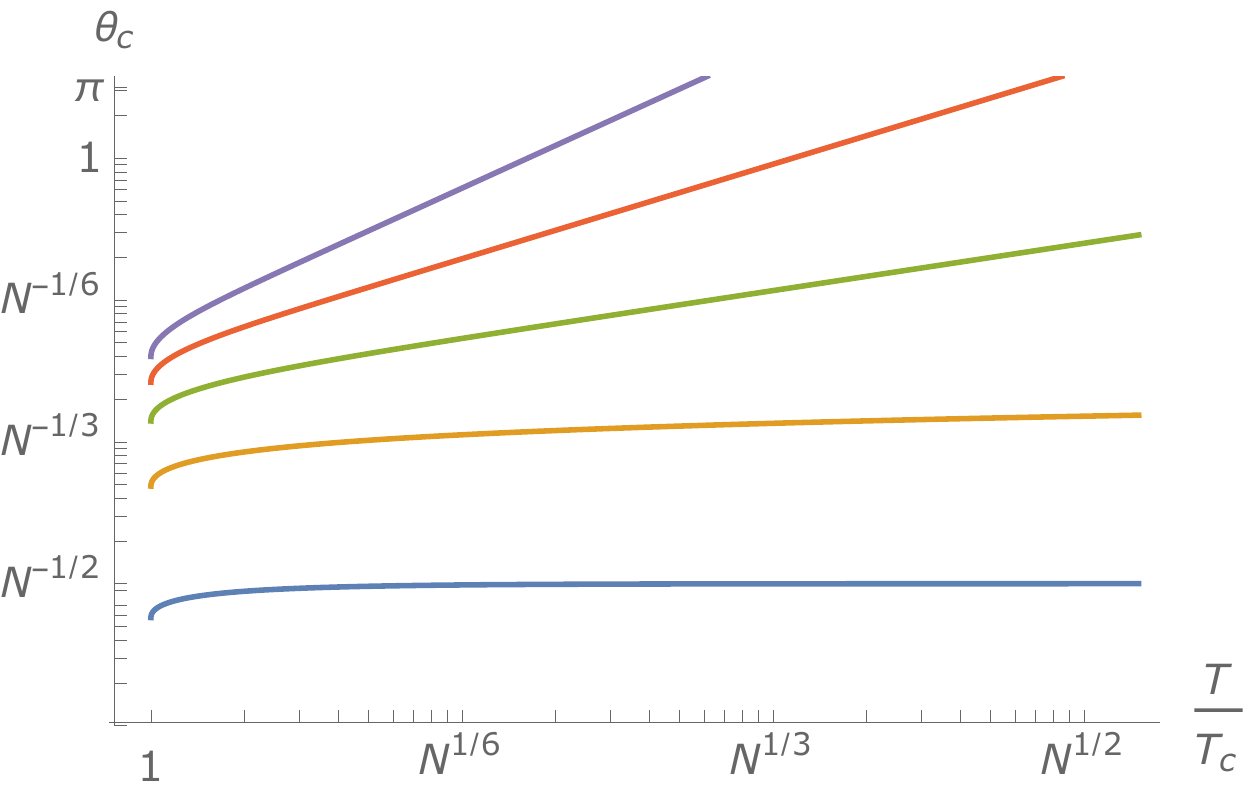}\hfill
		\includegraphics[width=.49\textwidth]{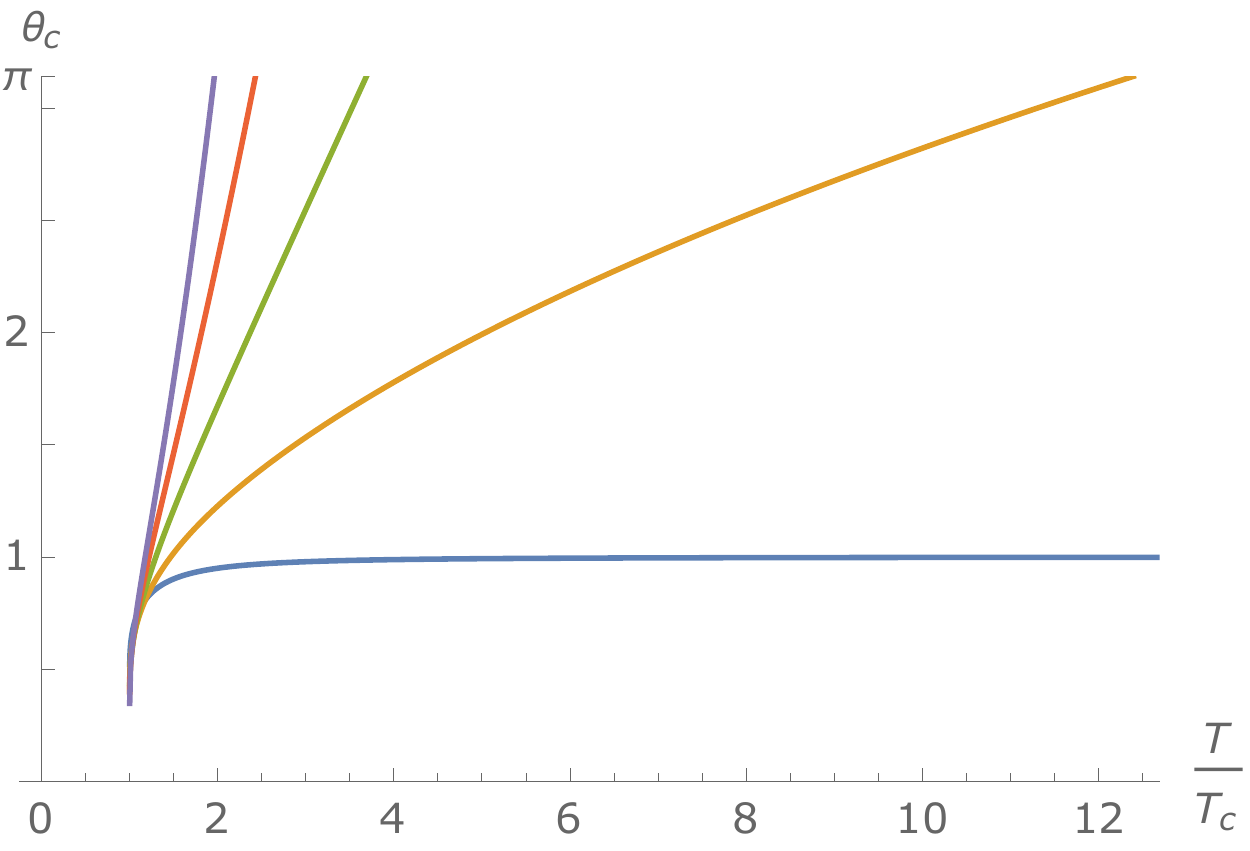}
		\caption{\label{fig:thetac}Crossover scale $\theta_c$ in vector (left) and adjoint (right) models as a function of temperature for dimensions $d=3$ to 7 (bottom to top).  
		}
	}
\end{figure}

We summarize our results in Table \ref{tab:comp2} and in Figures~\ref{fig:phasediagram}
to~\ref{fig:thetac}. At temperatures below the phase transition the leading order correlation 
function is given by a thermal boundary-to-boundary correlator in $d{+}1$-dimensional 
AdS spacetime, up to corrections suppressed by powers of $1/N$, in both models. 
This corresponds to the leftmost regions labelled by $I$ in Figures~\ref{fig:phasediagram} 
and~\ref{fig:phasediagramAdj}.  

Above the critical temperature the behavior 
of the correlation function is qualitatively different depending on whether the separation 
between the operators is larger or smaller than the thermal wavelength. 
At subthermal distances, {\it i.e.} in the region labelled $I$ below the green dashed 
lines in Figures~\ref{fig:phasediagram} and~\ref{fig:phasediagramAdj}, the 
correlation function is still well described by the boundary-to-boundary correlator of 
$d+1$-dimensional thermal AdS space, but with corrections parametrized by $\theta/\beta$ 
instead of $1/N$.

At distances beyond the thermal wavelength, on the other hand, the correlation functions
differ significantly from their low-temperature counterparts. In this regime, there is a further 
division into intermediate and long distances labelled by by~$II$ and~$I'$, respectively, in 
Figures~\ref{fig:phasediagram} and~\ref{fig:phasediagramAdj}. At intermediate distances
(region~$II$), below the crossover scale $\theta_c$ indicated by orange dotted lines in 
Figures ~\ref{fig:phasediagram} and~\ref{fig:phasediagramAdj}, there is an effective
dimensional reduction and the correlation function reduces to the small angle limit of 
the vacuum boundary-to-boundary propagator in $d$-dimensional AdS spacetime, 
with a $T^2$ factor in front. 
At long distances (region~$I'$), above the crossover scale, the correlation function 
goes back to the form of a thermal propagator in $d{+}1$-dimensional AdS, but with a 
temperature dependent normalization factor. The crossover scale depends on
both temperature and the number of dimensions as shown in Figure~\ref{fig:thetac}.

\begin{table*}
	\centering
	\begin{tabular}{|c | c| c| c | c| c | c |} 
		\hline 
		\multirow{4}{*}{
		} 
		& \multicolumn{2}{|c|}{\multirow{3}{*}{$T\leq T_c^a \sim 1$}} 
		& \multicolumn{2}{|c|}{\multirow{3}{*}{$T_c^a \ll T \leq T_c^f \sim \sqrt{N}$}} 
		& \multicolumn{2}{|c|}{\multirow{3}{*}{$T\gg T_c^f$}}\\
		& \multicolumn{2}{|c|}{} & \multicolumn{2}{|c|}{} & \multicolumn{2}{|c|}{} \\ [2ex] 
		\cline{2-7} 
		& Vec & Adj & Vec & Adj & Vec & Adj \\
		\hline\hline
		& \multicolumn{2}{|c|}{} & \multicolumn{2}{|c|}{} & \multicolumn{2}{|c|}{} \\ 
		$\theta\ll\beta$ & \multicolumn{2}{|c|}{\multirow{9}{*}{${\cal K}^{d+1}_\text{AdS}(\beta,\theta)$}} & \multicolumn{2}{|c|}{${\cal K}^{d+1}_\text{AdS}(\beta,\theta)$} &  \multicolumn{2}{|c|}{${\cal K}^{d+1}_\text{AdS}(\beta,\theta)$} \\ [4ex] 
		\cline{1-1} \cline{4-7}
		& \multicolumn{2}{|c|}{} & & & \multicolumn{2}{|c|}{} \\ 
		$\theta_c\gg\theta\gg\beta$
		& \multicolumn{2}{|c|}{} &\multirow{5}{*}{${\cal K}^{d+1}_\text{AdS}(\beta,\theta)$} & $\frac{F_d (\theta)}{\beta^2}$ &  \multicolumn{2}{|c|}{$\frac{F_d(\theta)}{\beta^2}$} \\ [4ex] 
		\cline{1-1} \cline{5-7}
		& \multicolumn{2}{|c|}{} & & & & \\ 
		$\theta\gg \theta_c \gg\beta\gg\beta'_c$ & \multicolumn{2}{|c|}{} &  & \specialcell{$f_\beta{\cal K}^{4}_\text{AdS}(\beta,\theta)$\\\hspace{0.2em} for $d=3$} & $f_\beta {\cal K}^{d+1}_\text{AdS}(\beta,\theta)$ & \specialcell{$f_\beta {\cal K}^{4}_\text{AdS}(\beta,\theta)$\\\hspace{0.2em} for $d=3$} \\[4ex]
		\hline
	\end{tabular}
	\caption{Leading order correlation functions in all angular regimes for the scalar fields both 
	in the adjoint and the fundamental representations of $U(N)$. The critical temperatures for 
	the adjoint and the vector models are $T_c^a$ and $T_c^f$, respectively,
	${\cal K}_\text{AdS}^d(\beta,\theta)$ is the boundary-to-boundary correlator 
	in $d$-dimensional thermal AdS space, $F_3 (\theta)=\log(\theta)^2$ and 
	$F_{d>3}(\theta)=\theta^{2 (3-d)}$. In the vector model $f_\beta \equiv \sum_p\rho_{|p|}$  
	and for the adjoint case $f_\beta \equiv \sum_p\rho_{|p|}^2$.
	}
	\label{tab:comp2}
\end{table*}

Focusing on the form of the correlators, the preceding summary can be given a bulk 
interpretation relying only on rather general features of the AdS/CFT correspondence, 
including that physics at long distances on the boundary corresponds to physics deep 
in the bulk. 
\begin{itemize}
\item In the regions labelled by I, the boundary correlation function matches to leading 
order a correlator induced by propagation through a bulk thermal AdS space. 
The geometry seems to be unaffected by the heat bath below the critical temperature 
or sufficiently close to the boundary. We see corrections to the correlation function 
at order $1/N$ and we expect back reaction on the geometry as well at that
order.
\item At sufficiently high temperature there is a dense region, labelled by II, 
which is deconfined in the sense that the singlet constraint is ineffective. 
In this region the boundary correlators behave like short distance  
correlators in a $(d{-}1)$-dimensional unconstrained theory. 
\item The deconfined character of the propagators above the phase transition is 
confirmed by evaluating the expectation value of (the square of) a Polyakov loop.
This is done for both the vector and the adjoint model in Appendix~\ref{appendixb}. 
The form of our correlation functions along with the presence of a non-vanishing 
Polyakov loop in the high temperature phase is consistent with a change in the 
emerging geometry away from thermal AdS to a black hole like 
configuration \cite{Furuuchi2005}.
\item The deconfined behavior becomes apparent at operator 
separation on the boundary of order the thermal wavelength, $\theta \sim \beta$. 
This implies that at the critical temperature the deconfined phase appears at the AdS scale 
in the adjoint model, independent of $N$, and grows with temperature. In the vector 
model, on the other hand, the deconfined 
phase appears on a super-AdS scale at large $N$ already at the critical temperature. 
\item We confirm the observation made in \cite{ShenkerYin2011} that there are no 
black holes at the AdS scale in higher spin gravity. Our results do suggest
the presence of black hole like objects but they are parametrically larger than AdS scale 
at large $N$. 
\item In low dimensions, and at large angles, the boundary correlator probes a core region 
labelled I'.  Surprisingly, the angular dependence is again that of a propagator through 
$d{+}1$-dimensional thermal AdS space, but with a temperature dependent overall
normalization. In this region the singlet constraint again plays a role. It is as if the 
boundary correlator again detects an effective AdS geometry. Thinking about the 
deconfined phase as signalling the presence a black hole, one is tempted to 
interpret this as the boundary correlator probing the second asymptotic region 
of an eternal black hole.
\end{itemize}

Our work can be extended in various ways. It would be desirable to have a deeper 
understanding of the deep core region I' and in particular whether it remains present
away from the free limit. Time dependent correlators 
may contain some signal of the evanescent modes found in \cite{JevickiYoon2016} in 
the limit of a flat $O(N)$ boundary theory. Time evolution 
could also shed light on possible precursors of positive Lyapunov exponents and the 
onset of chaos \cite{ShenkerStanford2014,DvaliFlassigGomezPritzelWintergerst2013,Perlmutter2016}, known to be present in Yang-Mills 
theories at finite coupling.

\begin{acknowledgments}
This work was supported in part by the Swedish Research Council through the 
Oskar Klein Centre and under 
contract 621-2014-5838, the Icelandic Research Fund grant 163422-051 and
the University of Iceland Research Fund.
\end{acknowledgments}

\begin{appendix}

\section{Explicit expressions in $d=3$ for the vector model}\label{sec:app3d}
In this appendix we specialize to $d=3$, which is the most relevant case for studying 
the duality between CFT and higher spin gravity, and focus mainly on the high temperature 
behavior. 
Using Eq.~\eqref{eq:zs_ht}, the eigenvalue distribution at $T_c>T \gg 1$ can be 
written as
\begin{equation}
\rho(\lambda) \to \frac{1}{2\pi}+\left(-\frac{\pi}{6} 
+ \frac{(|\lambda|-\pi)^2}{2\pi}\right)\frac{N_f}{N}\,T^{2}\,.
\label{highTrho}
\end{equation}
At the critical temperature,
\begin{equation}
T_c = \frac{\sqrt{3}}{\pi}\sqrt{\frac{N}{N_f}}\,,
\end{equation}
the eigenvalue distribution goes to zero at $\lambda=\pi$ and at
$T \geq T_c$ the eigenvalue density vanishes in a finite interval,
$\lambda_m \leq |\lambda|\leq \pi$, and can be written
\begin{equation}
\rho(\lambda) = 
\left\{\begin{array}{ccc}
\frac{3}{2\pi\gamma^2}
\left(\left(1 - \frac{|\lambda|}{\pi}\right)^2
-\left(1 - \frac{\lambda_m}{\pi}\right)^2 \right) 
&\quad \textrm{if} \quad&|\lambda|\leq \lambda_m\,, \\
\quad 0  & \quad\textrm{if}\quad &\lambda_m<|\lambda|  \,,
\end{array}\right.
\end{equation}
where $\gamma\equiv T_c/T$. The normalization condition Eq.~\eqref{eq:distnorm} 
forces $\lambda_m$, the angle at which the eigenvalue distribution goes to zero, 
to satisfy the following cubic equation
\begin{equation}
-2\left(\frac{\lambda_m}{\pi}\right)^3
+3\left(\frac{\lambda_m}{\pi}\right)^2
=\gamma^2 \,.
\label{eq:lambda}
\end{equation}
which  physical solution 
At temperatures above the phase transition, 
{\it i.e.} for $0<\gamma<1$, this equation has a single physical solution 
inside the range $0<\lambda_m<\pi$. The high-temperature limit amounts to
$\gamma\rightarrow 0$ and then the terms on the left hand side of the cubic 
equation must also go to zero, in which case we can drop the higher order term
and obtain
\begin{equation}
\label{eq:lambdamht}
\lambda_m \to \frac{\pi}{\sqrt{3}}\frac{T_c}{T}\,.
\end{equation} 
In other words, the distribution becomes $\delta$-like around $\lambda=0$ when $T \to \infty$.

The boundary-to-boundary correlator in thermal AdS$_4$ takes a particularly simple form,
\begin{equation}\label{eq:K4}
{\cal K}^{4}_\text{AdS}(\beta,\theta)=\frac{1}{32\pi^2}\sum_{n=-\infty}^{\infty} 
\frac{1}{\cosh{n\beta}-\cos{\theta}}=\frac{1}{16\pi^2}\frac{(\pi-\theta)}{\beta\sin\theta}\,.
\end{equation}
Below the phase transition, the large distance two-point function given by 
Eq.~\eqref{eq:4ptcorrlong} becomes
\begin{equation}
\exv{\text{Tr}\, |\phi(0)|^2\,\text{Tr}\,|\phi(y)|^2}
= \frac{1}{16\pi^2}\frac{\pi-\theta}{\beta\sin\theta}\left(1+\frac{2}{\gamma^2}\right)\,.
\end{equation}

The explicit expression for the high-$T$ Fourier moments of the eigenvalue distribution for 
$d = 3$ and $n\neq0$ reads
\begin{equation}
\label{eq:rhoHT}
\rho_{|n|} = 
\frac{6}{\pi^2 \gamma^2 n^2}\left(1+(\frac{\lambda_m}{\pi}-1) 
\cos \lambda_m n-\frac{\sin \lambda_m n}{\pi n}\right)\,,
\end{equation}
which is valid for $\gamma<1$ and where $\lambda_m$ is the physical solution of 
Eq.~\eqref{eq:lambda}. Note that by using this form of the distribution function we limit 
ourselves to the leading order behavior in $\beta$, independent of the distance regime 
under consideration.

In the short distance regime, $\theta\ll\beta$, we can explicitly compute the subleading 
contributions using Eq.~\eqref{eq:rhoHT}. The two-point function becomes
\begin{equation}
\exv{\text{Tr}\, |\phi(0)|^2\,\text{Tr}\,|\phi(y)|^2}
\approx\,\frac{1}{16\pi^2\theta^2}+\frac{3}{2\pi^5 \gamma^2 \beta\theta} 
\Bigg(\pi\zeta(3)+(\lambda_m-\pi) \Re Li_3 (e^{i \lambda_m})-\Im Li_4 (e^{i \lambda_m})\Bigg) \,.
\end{equation}
In the limit of very high temperature, $\gamma\ll1$, it further simplifies to
\begin{equation}
\exv{\text{Tr}\, |\phi(0)|^2\,\text{Tr}\,|\phi(y)|^2}  \approx  \frac{1}{16\pi^2\theta^2}
+\frac{1}{8\pi^2\beta\theta}  \Bigg(3+2 \log{\frac{\sqrt{3}}{\pi\gamma}}\Bigg)\,.
\end{equation}
In $d=3$, the crossover scale is $\theta_c\sim\beta_c\ll1$. Hence there is a clear 
distinction between intermediate and long distances. For the latter, $\theta\gg\theta_c$,  
and the correlator reads
\begin{align}
16\pi^2\exv{\text{Tr}\, |\phi(0)|^2\,\text{Tr}\,|\phi(y)|^2}
&\approx\,\frac{\pi-\theta}{\beta\sin\theta} 
\left(1+2\sum_{p = 1}^{\infty}\rho_{p}\right)\nonumber\\
&=\frac{\pi-\theta}{\beta\sin\theta} {\Bigg(1+\frac{2}{\gamma^2}
+\frac{12(\lambda_m-\pi)}{\pi^3\gamma^2} \Re Li_2 (e^{i \lambda_m})
-\frac{12}{\pi^3\gamma^2}\Im Li_3 (e^{i \lambda_m})\Bigg)} \nonumber\\
&=\frac{\pi-\theta}{\beta\sin\theta} \Bigg(\frac{2 \sqrt{3}}{\gamma}-1\bigg)\,,
\end{align}
where in the last expression the limit of very high temperature, 
$\lambda_m\sim \pi \gamma/\sqrt{3}$, is assumed and Eq.~\eqref{eq:K4} has been used.

For the intermediate regime, $\beta\ll\theta\ll\theta_c$, we can perform the sum in 
Eq.~\eqref{eq:id_fun}, leading to
\begin{equation}
\exv{\text{Tr}\, |\phi(0)|^2\,\text{Tr}\,|\phi(y)|^2} \approx\,\frac{1}{2\pi^2\beta^2}K^2\left(\cos^2 \frac{\theta}{2}\right)\,,
\end{equation}
where $K(x) =\int_{0}^{\pi/2} d\phi (1-x \sin^2\phi)^{-1/2}$ is the complete elliptic integral of the first kind. In the small angle approximation this yields
\begin{equation}
\exv{\text{Tr}\, |\phi(0)|^2\,\text{Tr}\,|\phi(y)|^2} \approx \,\frac{1}{2\pi^2\beta^2}\log^2\frac{\theta}{8}\,.
\end{equation}

\section{Polyakov loops}\label{appendixb}

In the adjoint model, the Polyakov loop itself vanishes due to the $\mathbb{Z}_n$ center symmetry of the action. As usual, we therefore consider the squared modulus of the loop, defined via 
\begin{equation}
	\exv{|{\cal P}|^2} = \frac{1}{N^2}\left\langle \text{Tr} \,e^{i\int_0^\beta dt A_0}\, \text{Tr} \,e^{-i\int_0^\beta dt A_0}\right\rangle
	= \left\langle\left|\int d\lambda\, \rho(\lambda)\, e^{i\lambda}\right|^2\right\rangle = \left\langle\rho_1^2\right\rangle\,.
\end{equation}
Below the phase transition, this vanishes to leading order in both the adjoint and the fundamental models. However, in the case of the fundamental model, it receives contributions at order $1/N$ 
due to corrections to the saddle point,
\begin{equation}
	{\cal P} = \sqrt{\exv{|{\cal P}|^2}} = \frac{N_f}{N}z^d_S(x)\,. 
\end{equation}
In the matrix model, on the other hand, no contributions to the Polyakov loop arise from the 
saddle point but contributions do arise at order $1/N^2$ from fluctuations around the saddle. 
We obtain them by inserting $\rho_1^2$ into the path integral Eq.~\eqref{eq:genfun}, 
\begin{equation}
	{\cal P} = \frac{1}{2 N^2 (1 - z^d_S(x))} \,.
\end{equation}

Above the phase transition, the (square of the) Polyakov loop picks up nonvanishing 
contributions in either model. In the vector model, it can be read from 
Eq.~\eqref{eq:asymfourier} and is found to interpolate between 
$\frac{1}{2(1-2^{2-d})\zeta(d-1)}$ at $T_c$ and $1$ for $T \to \infty$. 
Similarly, in the adjoint model, ${\cal P}$ interpolates between $1/2$ and $1$.

\end{appendix}

\bibliographystyle{utphys}
\bibliography{Bibliography}
\end{document}